\documentclass[aip,jap,reprint]{revtex4-1}
\usepackage[english]{babel}

\usepackage{amsmath}
\usepackage{graphicx}
\usepackage{natbib}
\usepackage{verbatim}
\usepackage[withbib, all]{authorindex}	
\usepackage[usenames]{color}
\usepackage{bm}
\usepackage[bookmarks=false]{hyperref}

\hypersetup{
    colorlinks, linkcolor={blue},
    citecolor={blue}, urlcolor={blue},
}
\begin{document}

\title{Magnetization dynamics of cobalt grown on graphene}

\author{A. J. Berger}
\affiliation{Department of Physics, The Ohio State University, Columbus, Ohio 43210, USA}

\author{W. Amamou}
\affiliation{Department of Physics and Astronomy, University of California, Riverside, CA 92521, USA}

\author{S. P. White}
\affiliation{Department of Physics, The Ohio State University, Columbus, Ohio 43210, USA}

\author{R. Adur}
\affiliation{Department of Physics, The Ohio State University, Columbus, Ohio 43210, USA}

\author{Y. Pu}
\affiliation{Department of Physics, The Ohio State University, Columbus, Ohio 43210, USA}

\author{R. K. Kawakami}
\affiliation{Department of Physics, The Ohio State University, Columbus, Ohio 43210, USA}
\affiliation{Department of Physics and Astronomy, University of California, Riverside, CA 92521, USA}

\author{P. C. Hammel}
\email{hammel@physics.osu.edu}
\affiliation{Department of Physics, The Ohio State University, Columbus, Ohio 43210, USA}

\date{\today}


\newpage
\begin{abstract}
Ferromagnetic resonance (FMR) spin pumping is a rapidly growing field which has demonstrated promising results in a variety of material systems. This technique utilizes the resonant precession of magnetization in a ferromagnet to inject spin into an adjacent non-magnetic material.  Spin pumping into graphene is attractive on account of its exceptional spin transport properties.  This article reports on FMR characterization of cobalt grown on CVD graphene and examines the validity of linewidth broadening as an indicator of spin pumping.  In comparison to cobalt samples without graphene, direct contact cobalt-on-graphene exhibits increased FMR linewidth---an often used signature of spin pumping. Similar results are obtained in Co/MgO/graphene structures, where a 1nm MgO layer acts as a tunnel barrier.  However, SQUID, MFM, and Kerr microscopy measurements demonstrate increased magnetic disorder in cobalt grown on graphene, perhaps due to changes in the growth process and an increase in defects. This magnetic disorder may account for the observed linewidth enhancement due to effects such as two-magnon scattering or mosaicity.  As such, it is not possible to conclude successful spin injection into graphene from FMR linewidth measurements alone.
\end{abstract}

\maketitle

Microwave-driven FMR has been shown to be a viable means for injecting spins into non-magnetic materials such as Pt, Au, GaAs, Si, and possibly graphene \cite{wang_large_2013, czeschka_scaling_2011, mosendz_quantifying_2010, ando_electrically_2011, ando_observation_2012, patra_dynamic_2012,tang_dynamically_2013}.  This is accomplished by placing a ferromagnet (FM) in contact with the non-magnetic material (NM), and driving the FM magnetization into precession.  The free carriers of the NM absorb some angular momentum from the precessing FM magnetization, thus becoming spin polarized.  This angular momentum transfer will enhance the damping of the FM.  Spin pumping serves as an alternative to electrical spin injection schemes in which a bias current carries spin-polarized charges across the FM/NM interface \cite{han_spin_2012, tombros_electronic_2007}.  The spin pumping effect was first observed by monitoring the FMR linewidth, a measure of the total damping of the FM magnetization by several available mechanisms \cite{urban_gilbert_2001, heinrich_dynamic_2003}.  However, the magnetization dynamics of a thin film FM are also extremely sensitive to sample defects and disorder.  Graphene's low surface energy and high surface diffusion cause the deposited metal atoms to form clusters and make growth of smooth layers difficult \cite{han_spin_2012, zhou_deposition_2010}.  As a consequence, we anticipate that the ferromagnetic properties of thin film cobalt and possibly other ferromagnets grown on graphene will be affected.  In this report, we reproduce the linewidth broadening of cobalt FMR spectra due to underlying graphene \cite{patra_dynamic_2012}.  Additional magnetostatic and dynamic characterizations make evident that substantial changes have been made to the cobalt films by growth on graphene.

\label{sec:Experiments}
Single layer graphene is synthesized by low pressure chemical vapor deposition (CVD) on Cu foil using a Cu enclosure \cite{li_large-area_2011}. The resulting large area graphene ($\sim5\mbox{ mm} \times 5$ mm) is then transferred to SiO$_2$(300 nm)/n-Si substrate by etching the Cu foil while using poly-methyl methacrylate (PMMA) for mechanical support. MgO and Co films are deposited at room temperature in a molecular beam epitaxy chamber with base pressure of $1 \times 10^{-10}$ torr. Co is deposited from a thermal effusion cell at a rate of $\sim$1.6 ${\mbox{\AA}}$/min. MgO is deposited from an e-beam source at a rate of $\sim$0.9 ${\mbox{\AA}}$/min. For direct contact samples (which we will denote Co/Gr), we deposit 15 nm of Co directly onto graphene. For tunnel barrier samples \cite{han_tunneling_2010} (Co/MgO/Gr), we deposit 1 nm of MgO onto graphene before growing the 15 nm Co film.  For both direct contact and tunnel barrier devices, we prepare control devices without graphene (Co and Co/MgO).  All samples are capped with 4 nm MgO to reduce oxidation of the Co.

Using a Bruker EMX-Plus EPR (electron paramagnetic resonance) system (with microwave cavity supporting the TE$_{102}$ mode resonant at 9.8 GHz), we measure the FMR response of Co for the various samples as a function of in-plane angle (orientation of the external static magnetic field $H_0$ with respect to the FM easy axis).  Following the numerical methods of Ref. \onlinecite{du_control_2013}, we determine the cobalt effective saturation magnetization ($4 \pi M_{\rm{eff}} = 4 \pi M_s - H_{\perp}$, where $H_{\perp}$ is the out-of-plane uniaxial anisotropy), in-plane cubic ($H_{\rm{cubic}}$), and in-plane uniaxial ($H_{\rm{uniaxial}}$) anisotropy fields (see Table \ref{tab:FMRProps}).  At this fixed frequency we observe an enhanced linewidth for samples with graphene, as compared to the controls, at all field angles.  The variation in $M_{\rm{eff}}$ and anisotropy as determined by these measurements, while larger for our samples than in Ref. \onlinecite{patra_dynamic_2012}, cannot account entirely for the linewidth enhancement.  We therefore turn to broadband FMR measurements.

\begin{figure}
	\includegraphics[width=.75\linewidth]{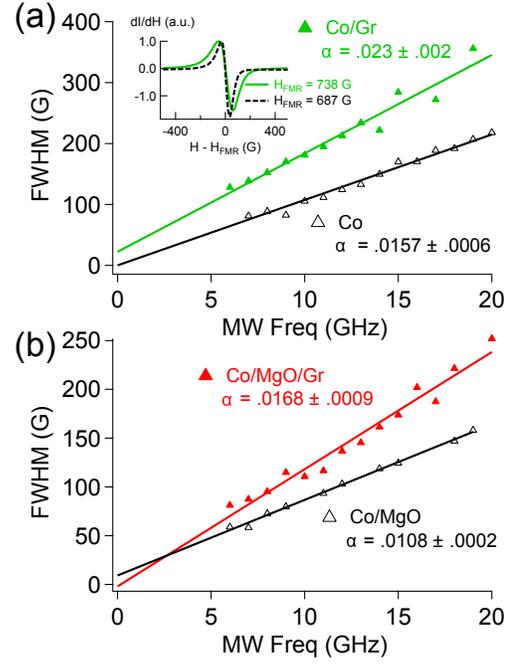}
	\caption{(a) FMR FWHM vs. MW frequency for Co/Gr (green closed triangles) and Co (black open triangles).  Inset: Example FMR spectra of Co/Gr (solid green) and Co (dashed black) samples at 10 GHz.  (b) FMR FWHM vs. MW frequency for tunnel barrier Co/MgO/Gr (red closed triangles) and Co/MgO (black open triangles).}
	\label{fig:HWHMvsFreq}
\end{figure}

\noindent
\begin{table}
\caption{Effective saturation magnetization and anisotropy fields extracted from in-plane rotation dependence of FMR.}
\begin{tabular}{ | l | l | l | l | }
		\hline
		Sample & $4 \pi M_{\rm{eff}}$ (G) & $H_{\rm{cubic}}$ (G) & $H_{\rm{uniaxial}}$ (G)    \\
		\hline
		\hline
	  Co/Gr & 15780 & 2 & 12 \\
		Co & 18180 & 0 & 21 \\
		Co/MgO/Gr & 17720 & 2 & 33 \\
		Co/MgO & 19370 & 2 & 23 \\
		\hline	
\end{tabular}
\label{tab:FMRProps}
\end{table}
Using a coplanar stripline, and the ``flip-chip'' method to place the Co film as near the microwave source as possible, we measure the FMR spectrum at various microwave frequencies between 5 and 20 GHz.  Because we use field modulation and lock-in techniques, the resulting microwave absorption vs. $H_0$ (Fig. \ref{fig:HWHMvsFreq} inset) is fit to a Lorentzian derivative lineshape.  The extracted widths (full width at half-maximum) are shown in Fig. \ref{fig:HWHMvsFreq}.  Such a curve is often fit to the dependence
\begin{equation}
\Delta H_{\rm{FWHM}} = \Delta H_{\rm{inhom}} + \frac{2 \alpha \omega}{\gamma}
\label{eq:linearLW}
\end{equation}
\noindent The frequency-independent linewidth (y-intercept), $\Delta H_{\rm{inhom}}$, is known as inhomogenous broadening and describes spatial variations in the $M_s$ or anisotropy fields which locally shift the resonant condition.  A bulk FMR measurement integrates over these inhomogeneous regions, resulting in a broadened line.  The second term, with gyromagnetic ratio $\gamma$ and microwave angular frequency $\omega$, is the linewidth contribution, $\Delta H_{\rm{Gilbert}}$, due to Gilbert damping of the material.  In the phenomenological Landau-Lifshitz-Gilbert equation describing the dynamics of a FM, the term $\alpha$ appears as a viscous damping term, such that at higher frequencies the damping increases.   

Spin pumping introduces additional Gilbert damping: $\alpha = \alpha_0 + \alpha_{\rm{sp}}$, where $\alpha_0$ represents the Gilbert damping intrinsic to the material and $\alpha_{\rm{sp}}$ is the addition due to spin pumping \cite{tserkovnyak_enhanced_2002}.  Our Co/Gr sample shows a $46\%$ enhancement in $\alpha$ extracted from Fig. \ref{fig:HWHMvsFreq} as compared to the Co control, while the tunnel barrier sample shows a $56\%$ enhancement over its corresponding control.  If this enhancement is attributed to spin pumping \cite{patra_dynamic_2012}, the resulting spin-mixing conductance \cite{mosendz_quantifying_2010} $g_{\uparrow\downarrow} = 4 \pi M_s t_{\rm{FM}} \alpha_{\rm{sp}} / (g \mu_{\rm{B}})$ is comparable to or larger than that found in Co/Pt spin pumping bilayers \cite{czeschka_scaling_2011}.  Such a result is surprising, given graphene's much longer spin lifetime and much lower carrier density as compared to Pt.

Mechanisms other than those included in Eq. \ref{eq:linearLW} can broaden the FMR line.  For example, disorder in the film give rise to two-magnon scattering ($\Delta H_{\rm{2mag}}$) and mosaic-like varation in the magnetocrystalline axes or other sample parameters ($\Delta H_{\rm{mosaic}}$) \cite{zakeri_spin_2007}.  We can experimentally address the role of two-magnon scattering as follows.  For magnetization in-plane, the uniform FMR mode can scatter off sample defects into same-frequency finite wavelength magnon modes.  Modes parallel and perpendicular to the magnetization have different frequency-wavevector dispersions, which create a band of available magnonic states for scattering \cite{sparks_ferromagnetic-relaxation_1964}. As the magnetization is rotated out-of-plane, this magnon band shifts up in frequency, eventually eliminating all $k \neq 0$ magnons at the uniform mode frequency \cite{hurben_theory_1998}.  As such, out-of-plane FMR can be measured to eliminate the two-magnon scattering linewidth contribution.   

\begin{figure}
	\includegraphics[width=0.75\linewidth]{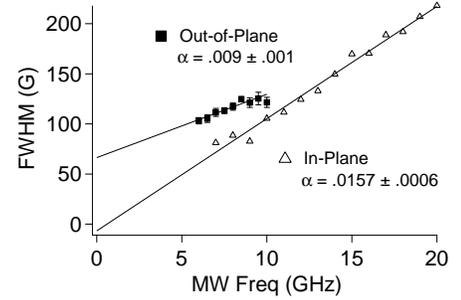}
	\caption{Comparison of 90deg out-of-plane (solid black squares) to field in-plane FMR linewidth (black open triangles) vs. frequency for cobalt control sample.  In-plane data are reproduced from Fig. \ref{fig:HWHMvsFreq} for comparison.  The large discrepancy in slope indicates that the in-plane linewidth includes a substantial two-magnon contribution.}
	\label{fig:OOPvsIP}
\end{figure}

Fig. \ref{fig:OOPvsIP} is consistent with two-magnon scattering in our thin-film Co samples---the slope is reduced in the out-of-plane configuration, implying that the in-plane measurements include a mixture of Gilbert and two-magnon damping.  We also note that the inhomogeneous broadening contribution can be larger for out-of-plane measurements, since the resonance condition varies linearly with $M_{\rm{eff}}$ for out-of-plane, but only as $\sqrt{M_{\rm{eff}}}$ for in-plane.  Previous studies \cite{lindner_non-gilbert-type_2003} have demonstrated the utility of performing in-plane measurements over a broader range of frequencies in order to properly disentangle the various linewidth contributions.  In the limited frequency range of 5-20 GHz, Gilbert damping cannot be distinguished from two-magnon scattering using only in-plane FMR measurements, and a linear extrapolation to obtain $\Delta H_{\rm{inhom}}$ can have large inaccuracy.  In order to further investigate sample disorder, we look into the magnetostatics of the various films.  

\begin{figure}
	\includegraphics[width=\linewidth]{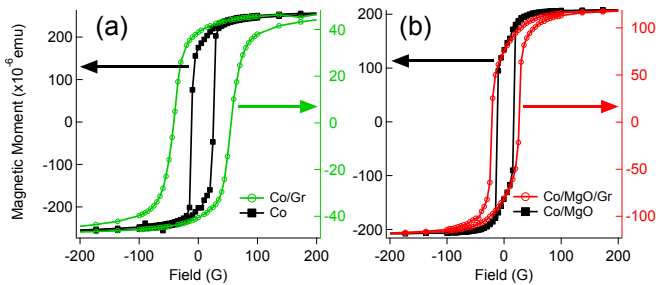}
	\caption{Magnetization curves obtained by SQUID magnetometer exhibiting increased coercivity for films grown on graphene. (a) Co/Gr (green open circles) and Co (black closed squares).  (b) Co/MgO/Gr (red open circles) and Co/MgO (black closed squares).}
	\label{fig:SQUID_MvsH}	
\end{figure}

Growth of metals on graphene will proceed very differently from growth on other substrates.  Because of the low metal-carbon bond energy, Co tends to form low density, large sized clusters during initial growth stages (sub-monolayer) \cite{zhou_deposition_2010}, promoting multi-grain films.  SQUID (superconducting quantum interference device) magnetometry data are consistent with this growth discrepancy, with samples grown on graphene displaying enhanced magnetic coercivity (Fig. \ref{fig:SQUID_MvsH}), likely due to an increase in domain wall pinning centers.

\begin{figure}
	\includegraphics[width=\linewidth]{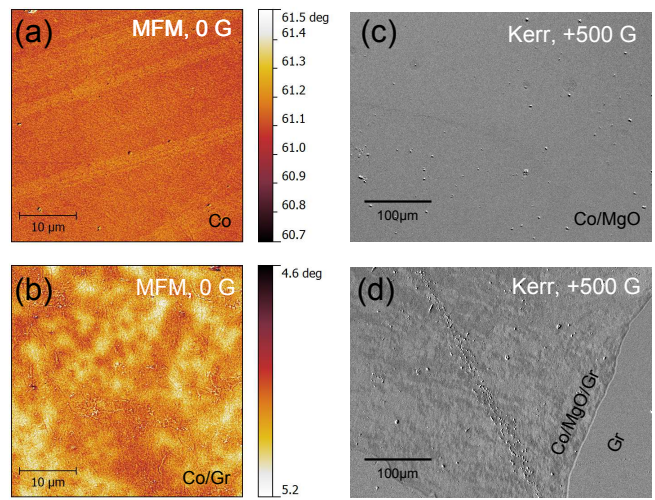}
	\caption{(a), (b) Comparison of MFM images for (a) Co and (b) Co/Gr samples at zero applied field.  The diagonal stripes in (a) are a topographic artifact.  (c), (d) Comparison of Kerr images for (c) Co/MgO control and (d) Co/MgO/Gr samples at 500G.}
	\label{fig:DomainImages}	
\end{figure}

%
To gain more direct information regarding the magnetic disorder in cobalt, we image the films with magnetic force and Kerr microscopies.  In both measurements, films grown on graphene (Fig. \ref{fig:DomainImages}b,d) exhibit magnetic variation that is absent in samples without graphene (Fig. \ref{fig:DomainImages}a,c).  Magnetic force microscopy (MFM) images of the Co/MgO/Gr sample (not shown) display magnetic nonuniformity similar to Fig. \ref{fig:DomainImages}b, while the Co/MgO control displays smooth magnetization.  Likewise, Kerr microscopy images of the Co/Gr sample (not shown) exhibit pronounced magnetic variation relative to the Co control, although it does not persist as prominently to high fields as in the tunnel barrier sample (Fig. \ref{fig:DomainImages}d).  Even though magnetic domains may be saturated by the $H_0$ necessary to achieve resonance for the FMR measurements, the defects responsible for domain wall pinning and the magnetic variation imaged in Fig. \ref{fig:DomainImages} can continue to act as scattering sites for two-magnon-type relaxation, or create variation in crystallite orientation in the cobalt film.  

Our magnetic characterization measurements indicate that inference of spin pumping from linewidth broadening is not simple.  Spin pumping is just one of many mechanisms which can broaden an FMR line.  Many characteristics of a FM can be altered by growth on graphene, including not only its Gilbert damping, but also anisotropy and additional damping mechanisms such as two-magnon scattering.  Direct spin transport measurements would improve confidence that FMR spin pumping is effective in graphene.

This research was supported by funding from the Center for Emergent Materials at the Ohio State University NSF MRSEC Award Number DMR-0820414 and by the US Department of Energy through award number DE-FG02-03ER46054 (R.A., S.W.).  Technical support was provided by the NanoSystems Laboratory at OSU.


%

\end{document}